\documentclass[twocolumn]{aastex631}
\usepackage[normalem]{ulem}  %\sout
\usepackage[T1]{fontenc}
\usepackage{ae,aecompl}
\usepackage{graphicx}
\usepackage{amsmath}
\usepackage{amssymb}
\usepackage{supertabular}
\usepackage{hyperref}
\usepackage{xcolor}
\usepackage{academicons}
\usepackage{multirow}
\usepackage{booktabs}

\begin{document}%--------|---------|---------|---------|---------|---------|---------|---------|
\title{Delayed Thermal Relaxation of Rapidly Cooling Neutron Stars: Nucleon Superfluidity and Non-nucleon Particles}
\author[0000-0001-6836-9339]{Zhonghao Tu}
\affiliation{Department of Astronomy, Xiamen University, Xiamen, Fujian 361005, China}
\author[0000-0001-9849-3656]{Ang Li}
\affiliation{Department of Astronomy, Xiamen University, Xiamen, Fujian 361005, China}

\correspondingauthor{Ang Li}
\email{liang@xmu.edu.cn}

\date{\today}
\begin{abstract}
The thermal relaxation time of neutron stars, typically defined by a sudden drop in surface temperature, is usually on the order of 10 to 100 years.
In this study, we investigate neutron star thermal relaxation by incorporating nucleon superfluidity and non-nucleonic particles, specifically considering hyperons as a representative case.
We find that rapidly cooling neutron stars driven by neutron superfluidity and direct Urca processes demonstrate delayed thermal relaxation under specific physical conditions. The former acquires that the neutron $^3P_2$ critical temperature is small enough, whereas the latter depends on the presence of a small core that permits direct Urca processes. To explore these scenarios, we propose simple theoretical frameworks to describe these delayed thermal relaxation behaviors and discuss how an recently-established enhanced modified Urca rate influences the relaxation time. By confronting the theoretical results with the observation of Cassiopeia A, we can effectively constrain the maximum neutron $^3P_2$ critical temperature.
\end{abstract}


%\section{Motivation}
\section{Introduction}%--------|---------|---------|---------|---------|---------|---------|---------|
The cooling of isolated neutron stars (NSs) can serve as a powerful tool for probing the internal structure of NSs. NS cooling simulations need to properly consider the equation of state (EoS), composition, and superfluidity of dense matter, as well as transport properties, e.g., thermal conductivity, specific heat, and neutrino emissivity \citep{Pethick1992_RMP64-1133,Page2004_ApJSupp155-623,Yakovlev2004_ARAA42-169,Yakovlev1999_PU42-737,Yakovlev2001_PR354-1,Yakovlev2015_MNRAS453-581,Potekhin2015_SSR191-239}. From the theoretical side, there are many uncertainties in the calculations of these microscopic physics. Confronting cooling simulations and observations can provide new insights for microscopic theories. Some studies have already used observations to constrain the EoS \citep{Newton2013_ApJL779-L4,AlvarezSalazar2018_ApP103-67}, composition \citep{Yakovlev2004_ASR33-523,Raduta2019_MNRAS487-2639}, and the critical temperature of superfluids \citep{Page2000_PRL85-2048,Page2009_ApJ707-1131,Page2011_PRL106-081101,Shternin2021_MNRAS506-709,Raduta2017_MNRAS475-4347}.

The differences in microscopic physics between different structures of a NS can lead to a thermal decoupling between them, further resulting in the formation of heat sinks and cold fronts.
%and further results in the formation of heat sink and cold front.
The arrival of the cold front at the surface of the star signals the formation of the star's thermal coupling \citep{Lattimer1994_ApJ425-802,Potekhin1997_A&A323-415,Gnedin2001_MNRAS324-725}. The neutrino emissivity is a typical physical input and can be used to distinguish between different cooling scenarios. The standard cooling scenario is treated as being dominated by the modified Urca (mUrca) processes \citep{Friman1979_ApJ232-541}; on this basis, the minimal cooling scenario considers the effects of superfluidity, especially the Cooper breaking and formation (PBF) process \citep{Page2004_ApJSupp155-623,Page2009_ApJ707-1131,Grigorian2018_NPA980-105}; the enhanced cooling scenario includes any direct Urca (dUrca) processes involving nucleons and non-nucleonic particles if apprear \citep{Pethick1992_RMP64-1133,Prakash1992_ApJ390-L77,Lattimer1991_PRL66-2701}. The neutrino emissions of the PBF and dUrca processes are stronger than those of the mUrca processes and therefore may cause a NS to exhibit rapid cooling characteristics.

The sudden drop in the surface temperature of a NS is used to define the thermal relaxation time. The typical value of the thermal relaxation time ranges from 10 to 100 years, depending on the cooling model \citep{Lattimer1994_ApJ425-802,Gnedin2001_MNRAS324-725}. Superfluidity can shorten the thermal relaxation time by a factor of four \citep{Gnedin2001_MNRAS324-725}. The thermal relaxation time shows an anti-correlation with the NS mass.
However, the delayed thermal relaxation can be observed when a NS has a small size core that allows the dUrca processes in the case of nucleonic matter and without superfluidity~\citep{Sales2020_A&A642-A42}. Indeed, cooling simulations that include complicated internal physics may hinder the relation between key physical parameters and thermal relaxation properties.
Here we conduct the study of thermal relaxation of NSs to cases that include superfluidity and non-nucleonic particles.
We hope to bridge microscopic physics and thermal relaxation of NSs through a simple theoretical framework.

In this work, we construct several unified microscopic EoSs within the relativistic mean field (RMF) framework by using the effective interactions in different isospin vector and scalar channels. Taking full EoS thermodynamics as inputs, we perform cooling simulations for NSs with or without strangeness-bearing hyperons and investigate their thermal relaxation properties. We find that the rapid cooling NSs driven by the PBF and dUrca processes exhibit delayed thermal relaxation.

This paper is organized as follows. In Sec.~\ref{sec:theory_NSCool}, the theoretical framework for simulating NS cooling is given. Sec.~\ref{sec:theory_EoS}--\ref{sec:theory_sf} introduce the unified EoSs and superfluid models that we use for the cooling simulations, respectively.
In Sec.~\ref{sec:results}, we demonstrate that the delayed thermal relaxation can be observed considering nucleon superfluidity and non-nucleonic particles.
Sec. \ref{sec:relaxSF}--\ref{sec:relaxdUrca} are devoted to discussing why delayed thermal relaxation occurs in rapid cooling NSs. Two physcial models are proposed in Sec. \ref{sec:relaxSF}--\ref{sec:relaxdUrca}, with their corresponding analytical formulas also provided. Finally, a brief summary is given in Sec.~\ref{sec:summary}.

\section{Neutron Star Cooling and Physics Input}\label{sec:theory}
\subsection{Cooling of Neutron Stars}\label{sec:theory_NSCool}

The stellar cooling is described by the local energy balance and heat transport equations \citep{2006NuPhA.777..497P,1996NuPhA.605..531S},
\begin{align}
    & c_{v}\frac{\partial(Te^{\phi})}{\partial t}=-e^{2\phi}q_{\nu}-\frac{1}{4\pi r^2(1+z)}\frac{\partial(Le^{2\phi})}{\partial r}, \label{eq:energybalance} \\
    & Le^{2\phi} = -\frac{4\pi r^2\kappa e^{\phi}}{1+z}\frac{\partial(Te^{\phi})}{\partial r}, \label{eq:heattransport}
\end{align}
for relativistic stars.
Here, $T$ and $L$ are the stellar internal temperature and luminosity. $\kappa$, $c_{v}$ and $q_{\nu}$ are the local conductivity, specific heat and neutrino emissivity, respectively. $\phi$ is the metric function of the star, $1+z$ represents the gravitational red shift $1+z=(1-2m/r)^{-1/2}$ with the enclosed mass $m$ within the radial distance $r$. The inner and outer boundary condition for $L$ is $L(r=0)=0$ and $T_{\mathrm{b}} = T_{\mathrm{b}}(L_{\mathrm{b}})$; the location of outer boundary in the latter is defined such that $L_{\mathrm{b}}$ equals to the total photon luminosity of the star, i.e., $L_{\mathrm{b}}=4\pi R^2\sigma_{\mathrm{SB}}T_{\mathrm{e}}^4$, here $R$ is the NS radius, $T_{\mathrm{e}}$ is the effective surface temperature and $\sigma_{\mathrm{SB}}$ is the Stefan-Boltzmann constant. The relation between $T_{\mathrm{b}}$ and $T_{\mathrm{e}}$, so called ``$T_{\mathrm{e}}$-$T_{\mathrm{b}}$ relationship'', is applied with Fe envelope. For comparison with observations, we present the effective surface temperature at infinity, $T_{\mathrm{e}}^{\infty}=T_{\mathrm{e}}e^{\phi(R)}$, then the measurable luminosity $L^{\infty}$ at infinity can be obtained by $L^{\infty}=4\pi R^{\infty 2}\sigma_{\mathrm{SB}}T_{\mathrm{e}}^{\infty 4}$ with the radiation radius $R^{\infty}=Re^{\phi(R)}$.

From Eqs. (\ref{eq:energybalance}) and (\ref{eq:heattransport}), the NS cooling depends on both bulk and thermal dynamical properties of stellar matter. The global properties of the star are obtained by solving the Tolman--Oppenheimer--Volkoff (TOV) equation~\citep{Tolman1939_PR055-364,Oppenheimer1939_PR055-374},
\begin{equation}\label{equ:TOV}
\begin{aligned}
    \frac{\mathrm{d}P}{\mathrm{d}r} &= -\frac{\left[ P(r)+\varepsilon(r) \right]\left[ m(r)+4\pi r^{3}P(r) \right]}{r\left[ r-2m(r) \right]}, \\
    \frac{\mathrm{d}m}{\mathrm{d}r} &= 4\pi r^{2}\varepsilon(r),
\end{aligned}
\end{equation}
with the EoS, i.e., the pressure $P$ as a function of the energy density $\varepsilon$, as an input.
The thermal dynamical properties of the stellar matter, e.g., conductivity and neutrino emissivity, can be calculated with EoS and compositions of the star, we refer the details to \cite{Page2004_ApJSupp155-623}. Note that the dUrca processes are not permitted for all EoSs and compositions. The dUrca processes require that the EoS is stiff enough so that the proton fraction exceeds $\sim1/8$, or the possible presence of non-nucleonic particles in the NS cores.
Besides, superfluidity can play an important role in the cooling of NSs. On the one hand, superfluidity suppresses both the neutrino emission and specific heat; on the other hand, superfluidity opens a new rapid cooling mechanism, i.e., the PBF processes. In the following, we firstly construct the EoSs and the corresponding compositions, and then introduce the superfluid model used in this work.

\subsection{Equation of State}\label{sec:theory_EoS}
We adopt one model of quantum hadrodynamics \citep{Fetter1972_PT25-54,Walecka1974_APNY83-491,Serot1992_RPP55-1855}, e.g., relativistic mean field (RMF) model, to construct the EoSs for NSs. We consider the baryon octet interacting with each other through the exchange of isoscalar scalar and vector mesons ($\sigma$ and $\omega$), isovector vector meson ($\rho$) in the RMF model. The hidden-strangeness mesons ($\sigma^{\star}$ and $\phi$) are introduced to mediate the interaction between hyperons under the SU(6) symmetry. The Lagrangian density that describes the systems with time-reversal symmetry can be written as:
\begin{widetext}
\begin{equation}\label{equ:RMF_Lagrangian}
\begin{aligned}
    \mathcal{L} = & \sum_{B}\bar{\psi}_{B}\left\{ \gamma^{\mu}\left[ i\partial_{\mu}-g_{\omega B}\omega_{\mu}
                 -g_{\rho B}\boldsymbol{\rho}_{\mu}\boldsymbol{\tau}_{B}-g_{\phi B}\phi-q_BA_{\mu} \right]
                 -\left[ M_{B}-g_{\sigma B}\sigma-g_{\sigma^{\star}}\sigma^{\star}\right] \right\}\psi_{B} \\
                &+\frac{1}{2}(\partial^{\mu}\sigma\partial_{\mu}\sigma-m_{\sigma}^{2}\sigma^{2})+\frac{1}{2}(\partial^{\mu}\sigma^{\star}\partial_{\mu}\sigma^{\star}-m_{\sigma^{\star}}^{2}\sigma^{\star2})+U(\sigma,\omega_\mu,\boldsymbol{\rho}_{\mu})\\
                &-\frac{1}{4}W^{\mu\nu}W_{\mu\nu}+\frac{1}{2}m_{\omega}^{2}\omega^{\mu}\omega_{\mu}
                 -\frac{1}{4}\Phi^{\mu\nu}\Phi_{\mu\nu}+\frac{1}{2}m_{\phi}^{2}\phi^{\mu}\phi_{\mu}
                 -\frac{1}{4}\boldsymbol{R}^{\mu\nu}\boldsymbol{R}_{\mu\nu}+\frac{1}{2}m_{\rho}^{2}\boldsymbol{\rho}^{\mu}\boldsymbol{\rho}_{\mu}-\frac{1}{4}A^{\mu\nu}A_{\mu\nu}\\
                &+\sum_{l=e,\mu}\bar{\psi}_{l}(i\gamma_{\mu}\partial^{\mu}-m_{l}+e\gamma^0A_\mu)\psi_{l},
\end{aligned}
\end{equation}
\end{widetext}
where~$\boldsymbol{\tau}_{B}$ and $q_B$ are the Pauli matrices and charges of baryons, and~$M_{B}$~and~$m_{l}$~represent the baryon and lepton masses, respectively. $g_{mB}$ is the coupling constant between the meson $m$ and the baryon $B$. $\psi_{B(l)}$~is the Dirac field of the baryons or the leptons. $\sigma$, $\omega_{\mu}$, $\boldsymbol{\rho}_{\mu}$, $\sigma^{\star}$, and $\phi_{\mu}$ denote the quantum fields of mesons. The field tensors of $\omega$, $\rho$, $\phi$, and photon are
%%%%%%%%%%%%%%%%%%%%%%%%%%%%%%%%%%%%%%%%%%%%%%%%%%%%%%%%%%%%%%
\begin{equation}\label{equ:fieldtensor}
\begin{aligned}
    W_{\mu\nu}       &= \partial_{\mu}\omega_{\nu}-\partial_{\nu}\omega_{\mu}, \\
    \boldsymbol{R}_{\mu\nu} &= \partial_{\mu}\vec{\rho}_{\nu}-\partial_{\nu}\vec{\rho}_{\mu}, \\
    \Phi_{\mu\nu}       &= \partial_{\mu}\phi_{\nu}-\partial_{\nu}\phi_{\mu}, \\
    A_{\mu\nu} &= \partial_{\mu}A_{\nu}-\partial_{\nu}A_{\mu}.
\end{aligned}
\end{equation}
For reducing the additional degrees of freedom introduced by the manual matching of the crust and core EoSs, the approach for calculating a unified EoS using the RMF model has been developed. We briefly introduce below the methodology for constructing unified EoS, please refer to our previous work \citep{2024arXiv241209219T} for more details.

Using the mean-field approximation, the equations of motion of various mesons are obtained via the Euler-Lagrange equation,
\begin{align}
    (-\nabla^2+m_{\sigma}^{2})\sigma &= \sum_{B}g_{\sigma B}\rho_{s}^{B}+\frac{\partial U}{\partial\sigma}, \label{eom_1}\\
    (-\nabla^2+m_{\omega}^{2})\omega &= \sum_{B}g_{\omega B}\rho_{v}^{B}-\frac{\partial U}{\partial\omega}, \label{eom_2}\\
    (-\nabla^2+m_{\rho}^{2})\rho     &= \sum_{B}g_{\rho B}\rho_{v}^{B}\tau_{B}^{3}-\frac{\partial U}{\partial\rho}, \label{eom_3}\\
    (-\nabla^2+m_{\sigma^{\star}}^{2})\sigma^{\star} &= \sum_{B}g_{\sigma^{\star}B}\rho_{s}^{B}, \label{eom_4}\\
    (-\nabla^2+m_{\phi}^{2})\phi     &= \sum_{B}g_{\phi B}\rho_{v}^{B}, \label{eom_5}\\
    -\nabla^2A_0 &= e\left (q_B\sum_{B}\rho_{v}^{B}+q_l\sum_{l}\rho_{v}^{l} \right) , \label{eom_6}
\end{align}
where the meson fields have been replaced by their mean values. $\rho_v^{B(l)}$ and $q_{B(l)}$ are the vector density and charge of the baryon (lepton) species $B(l)$, respectively. $\rho_s^B$ and $\tau_B^3$ are the scalar density and isospin projection of the baryon species $B$, respectively. We assume that $\sigma^{\star}$ and $\phi$ (under SU(6) symmetry) mesons only couple to hyperons.

In the core of a NS, the translation invariance of the system eliminates the derivative terms in the Eqs. (\ref{eom_1})--(\ref{eom_6}). Then the particle fractions and meson fields can be obtained by iteratively solving the coupled nonlinear system composed of Eqs. (\ref{eom_1})--(\ref{eom_5}), beta equilibrium, charge neutrality, and baryon number conservation conditions. Furthermore, the energy density $\varepsilon_{\rm{core}}$ and pressure $P_{\rm{core}}$ can be derived from the energy-momentum tensor $\mathcal{T}^{\mu\nu}$ via $\varepsilon_{\rm{core}}=\langle\mathcal{T}^{00}\rangle$ and $P_{\rm{core}}=\langle\mathcal{T}^{kk}\rangle/3$, respectively.

For the crust of a NS, Eqs. (\ref{eom_4}) and (\ref{eom_5}) are not considered because the onset densities of hyperons are much higher than the crust-core transition density. Using the Wigner-Seitz and Thomas-Fermi approximations, along with the reflective boundary condition, we obtain the distributions of meson fields and particle densities by iteratively solving the coupled nonlinear system composed of Eqs. (\ref{eom_1})--(\ref{eom_3}) and (\ref{eom_6}), also including the same equilibrium conditions as in the core, in different geometries. In the outer crust, only spherical nuclei are considered. The nuclear pasta may emerge in the bottom of the inner crust, we consider five representative pasta geometries in the inner crust: droplets, rods, slabs, tubes, and bubbles. The ground state of the crust at a given density is obtained by determining the lowest energy among five pasta structures. By integrating the energy density functional for the optimal internal structure of the crust, we can obtain the energy density $\varepsilon_{\rm{crust}}$; then the pressure is evaluated using the thermodynamic relation, $P_{\rm{crust}}=\sum_{i=n,p,e} \mu_{i} \rho_{v}^{i} - \varepsilon_{\rm{crust}} $, with the chemical potentials of nucleons and electrons $\mu_i$.

\begin{table}[htbp]
  \centering
  \caption{Saturation properties of nuclear matter for original DD-ME2 and NL3. The saturation properties we list below include the saturation density~$\rho_{0}$~(fm$^{-3}$), binding energy per particle $E/A$ (MeV), incompressibility $K_0$ (MeV), skewness $Q_0$ (MeV), symmetry energy $J_0$ (MeV), slope of symmetry energy $L_0$ (MeV), and effective mass of neutron $M_{n}^{\star}/M_{n}$.}\label{tab:satpros}
  \setlength\tabcolsep{3pt}
  \begin{tabular}{lccccccc}
  \hline
  \hline
         & $\rho_{0}$ & $E/A$ & $K_0$ & $Q_0$ & $J_0$ & $L_0$ & $M_n^{\star}/M_n$ \\
                          & $\rm{fm}^{-3}$ & MeV & MeV & MeV & MeV & MeV \\
  \hline
    DD-ME2 & 0.152 & -16.14 & 251.1 & 479 & 32.30 & 51.26 & 0.572 \\
    NL3    & 0.148 & -16.24 & 272.2 & 198 & 37.4 & 118.5 & 0.594 \\
  \hline
  \end{tabular}
\end{table}

\begin{table}[htbp]
  \centering
  \caption{The coupling parameters $g_{\rho}$ and $a_{\rho}$ between nucleons and $\rho$ meson for different symmetry energy slope. $L_0=51.3$ MeV for original DD-ME2 and $L_0 = 118.5$ MeV for original NL3. These parameters are obtained by fixing the symmetry energy $E_{\rm{sym}}$ at $\rho_{\mathrm{B}}=0.11$ fm$^{-3}$ but adjusting the symmetry energy slope $L_0$ at the saturation density.}\label{tab:vectorcouple}
  \setlength\tabcolsep{3pt}
  \begin{tabular}{lcccccc}
    \hline
    \hline
    $L_0$ (MeV) & \multicolumn{2}{c}{original} & \multicolumn{2}{c}{60} & \multicolumn{2}{c}{80} \\
    \cmidrule(r){2-3} \cmidrule(r){4-5} \cmidrule(r){6-7}
      & $g_{\rho}$ & $a_{\rho}$ & $g_{\rho}$ & $a_{\rho}$ & $g_{\rho}$ & $a_{\rho}$ \\
    \hline
    DD-ME2 & 3.6836 & 0.5647 & 3.7917 & 0.4599 & 4.0097 & 0.2576 \\
    NL3    & 4.4744 & 0.0000 & 3.9359 & 0.4971 & 4.3241 & 0.1323 \\
    \hline
  \end{tabular}
\end{table}

\begin{table}%[htbp]
  \centering
  \caption{The transition density of EoS and global properties of NSs for different RMF effective interactions. The transition density of EoSs we list below include the outer-inner crust transition density $\rho_{\rm{oi}}$ and the crust-core transition density $\rho_{\rm{cc}}$. The listed global properties of NSs include:  the maximum mass $M_{\mathrm{TOV}}^{N}$ and the threshold mass $M_{c,N}^{np}$ which $np$ process in the case of without $\Lambda$ hyperons; the maximum mass $M_{\mathrm{TOV}}^{\Lambda}$, the threshold mass $M_{c,\Lambda}^{np}$ which $np$ process is active, and the threshold mass $M_{c,\Lambda}^{\Lambda p}$ which $\Lambda p$ process is active in the case of with $\Lambda$ hyperons.
  }\label{tab:EoSandNSproperties}
  \setlength\tabcolsep{3pt}
  \begin{tabular}{lcccccc}
    \hline
    \hline
     & \multicolumn{3}{c}{DD-ME2} & \multicolumn{3}{c}{NL3} \\
     \cmidrule(r){2-4} \cmidrule(r){5-7}
     $L_0$ (MeV) & 51.3 & 60 & 80 & 60 & 80 & 118.5 \\
    \hline
    $\rho_{\rm{oi}}$ ($10^{-4}$ fm$^{-3}$) & 1.994 & 2.035 & 2.129 & 2.038 & 2.101 & 2.228 \\
    $\rho_{\rm{cc}}$ (fm$^{-3}$) & 0.075 & 0.067 & 0.057 & 0.077 & 0.067 & 0.057 \\
    $M_{\rm{TOV}}^{N}$ ($M_{\odot}$) & 2.483 & 2.477 & 2.468 & 2.746 & 2.738 & 2.775 \\
    $M_{c,N}^{np}$ ($M_{\odot}$) & -- & -- & 1.552 & -- & 1.444 & 0.825 \\
    $M_{\rm{TOV}}^{\Lambda}$ ($M_{\odot}$) & 2.108 & 2.100 & 2.081 & 2.300 & 2.282 & 2.259 \\
    $M_{c,\Lambda}^{np}$ ($M_{\odot}$) & -- & -- & 1.966 & 2.239 & 1.445 & 0.825 \\
    $M_{c,\Lambda}^{\Lambda p}$ ($M_{\odot}$) & 1.309 & 1.294 & 1.281 & 1.446 & 1.4270 & 1.472 \\
    \hline
  \end{tabular}
\end{table}

\begin{figure}
\centering
\includegraphics[width=1.0\linewidth]{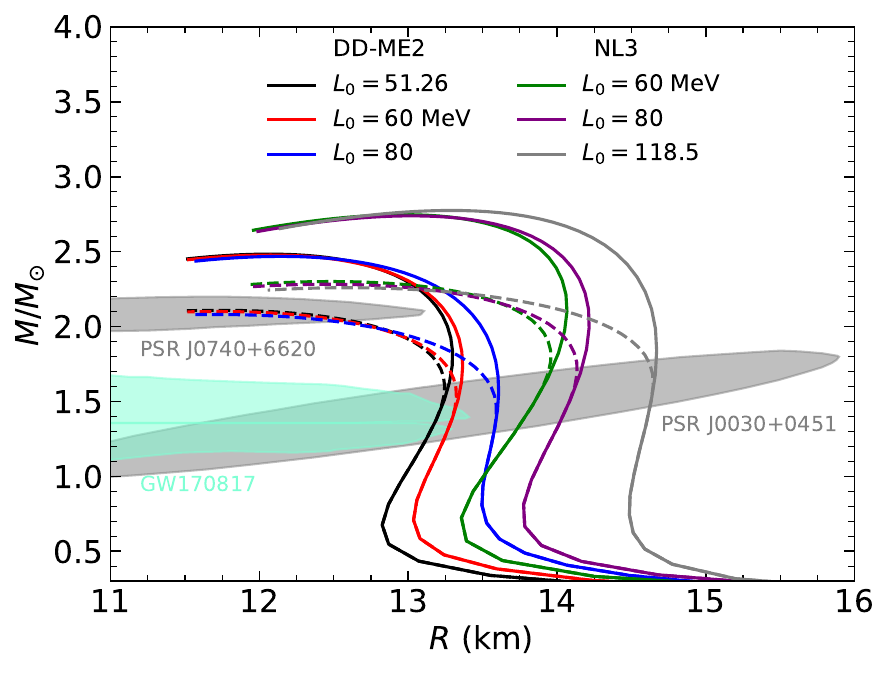}
\caption{\label{fig:MR}$M$-$R$ relations calculated with the unified DD-ME2 and NL3 EoS models for three choices of symmetry energy slope $L_0$ are shown in solid curves, with the corresponding results with the inclusion of hyperons shown in dashed curves. The mass-radius measurements from the NICER mission for PSRs J0030+0451 \citep{Vinciguerra2024_ApJ961-62} and J0740+6620 \citep{Salmi2024_ApJ974-294} are displayed. The mass-radius inferred from the GW170817 tidal deformability measurement are included \citep{Abbott2017_PRL119-161101}.
All these measurements are presented at the 90\% confidence level.
}
\end{figure}

In this work, we calculate the EoS using two sets of effective interaction: DD-ME2 and NL3. The original DD-ME2 and NL3 demonstrate support for the existence of massive NSs ($>2M_\odot$) with or without considering hyperons. By adjusting the density-dependent coupling, i.e., $g_{\rho}$ and $a_{\rho}$, for the $\rho$ meson, the effective interactions DD-ME2 and NL3 with the symmetry energy slope $L_0=$60 and 80 MeV are obtained, see \cite{Li2019_PRC100-015809} and \cite{Wu2021_PRC104-015802}. In Table. \ref{tab:satpros}, we can find that the characteristic coefficients of nuclear matter for the original effective interactions. The extensions of two original effective interactions in the isovector channel are listed in Table. \ref{tab:vectorcouple}. $K_0$ controls the behaviors of EoSs in the high density range, and $K_0$ of NL3 is larger than that of DD-ME2, this means that NL3 can produce heavier NS, as we can see in the $M$-$R$ relations of Fig. \ref{fig:MR}.
{$L_0$ strongly affects the EoS in the medium-density range and, consequently, governs the radius of a typical NS. The radius of a typical NS increases with increasing $L_0$, as shown in Fig. \ref{fig:MR}.}
As for unified EoS, the transition density $\rho_{\mathrm{oi}}$ between the outer crust and inner crust, and $\rho_{\mathrm{cc}}$ between the crust and core are also affected by $K_0$ and $L_0$, see Table. \ref{tab:EoSandNSproperties}. This reflect that the microscopic inputs used in our cooling simulations are more self-consistent.
We mention here that the selection of these effective interactions facilitates a comprehensive investigation of the dependence of NS thermal relaxation on the effective interaction in the isoscalar and isovector channels.

Hyperons emerge at high density range due to the fact that they are energetically more favorable than nucleons, leading to a drastically softening of the EoSs \citep{Sun2023_ApJ942-55,Ding2025_PRC111-014301}.
In the following sections, we refer to the NS that contains hyperon components as the hyperon star. Due to the repulsive interaction of $\Sigma$ hyperons \citep{Schaffner-Bielich2000_PRC62-034311,Wang2010_PRC81-025801} and the large mass of $\Xi^0$ hyperons, they appear at relatively large densities and their phase space is reduced \citep{Tu2022_ApJ925-16}, leading to insubstantial neutrino luminosity; the onset density of $\Xi^-$ hyperons is close to that of $\Lambda$ hyperons, while the pairing gap for $\Xi^-$ hyperons is enough large so that the dUrca processes involve $\Xi^{-}$ hyperons are strongly suppressed \citep{Raduta2017_MNRAS475-4347}. Based on the above considerations, only $\Lambda$ hyperons are taken into account in this work. Therefore, only the $np$ and $\Lambda p$ processes are possible dUrca processes in our calculations. In the framework of RMF model, the $\Lambda$-$N$ and $\Lambda$-$\Lambda$ interactions are determined by fitting the $\Lambda$ potentials in nuclear matter: $U_{\Lambda}^{(N)}(\rho_0)=-30$ MeV and $U_{\Lambda}^{(\Lambda)}(\rho_0/2)=-5$ MeV, where $\rho_0$ is the nuclear saturation density. The hyperon dUrca processes are possible because their conservation of momentum for three participating particles are satisfied easily \citep{Prakash1992_ApJ390-L77}.

From Table. \ref{tab:EoSandNSproperties}, the threshold mass $M_{c,\Lambda}^{\Lambda p}$ at which $\Lambda p$ process is activated in hyperon star is not significantly dependent on the EoS. For hyperon stars with masses exceeding $1.3M_{\odot}$ (DD-ME2) or $1.4M_{\odot}$ (NL3), $\Lambda p$ process is working. However, $np$ process is strongly dependent on the EoS. For DD-ME2, both for the NS and hyperon star, except for $L_0=80$ MeV, the $np$ process does not work inside the star below the maximum mass $M_{\mathrm{TOV}}$. NL3 produces a stiffer EoS, and thus the threshold mass $M_{c,N}^{np}$ or $M_{c,\Lambda}^{np}$ at which $np$ process is activated can be found below $M_{\mathrm{TOV}}$. A larger $L_0$ corresponds to a lower threshold mass. The current astrophysical observations have constrained $L_0$ to below 60 MeV \citep{Hooker2013_JPCS420-012153,Newton2013_ApJL779-L4,Tu2024_arXiv2412.09219}, the rapid cooling owing to the dUrca process is more likely observed in hyperon stars.

\subsection{Superfluidity}\label{sec:theory_sf}
\begin{figure}
\centering
\includegraphics[width=1.0\linewidth]{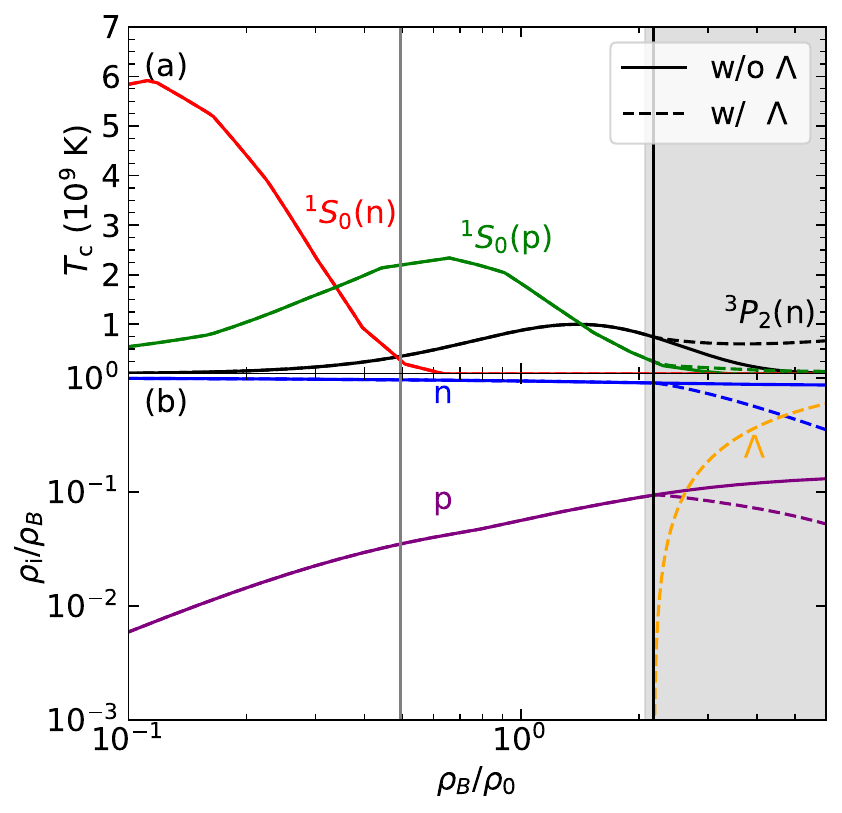}
\caption{\label{fig:DD-ME2_Tc_profile}Panel (a): The critical temperatures of neutron $^1S_0$ (red), proton $^1S_0$ (green), and neutron $^3P_2$ (magenta) as a function of the number density $\rho_B$ of NS matter, normalized to the nuclear saturation density. Panel (b): The baryon compositions of NS matter. The adopted effective interaction is the original DD-ME2 with $L_0=51.26$ MeV. The solid and dashed lines correspond to scenarios with and without $\Lambda$ hyperons, respectively. The gray-shaded region covers the range of central densities of NSs heavier than 1.2 $M_{\odot}$. The vertical black and gray lines represent the onset density of $\Lambda$ hyperons and the core-crust transition density of NSs, respectively. All critical temperatures in the figure are calculated for uniform NS matter. In actual cooling simulations, the critical temperatures below the core-crust transition density are determined by the crust model.
}
\end{figure}

Neutrons have $^1S_0$ pairing gap in the crust and $^3P_2$ pairing gap in the core. Proton $^1S_0$ pairing gap appears in the core. The paired nucleons enhance the neutrino emissivity by PBF processes in the NS core \citep{Page2004_ApJSupp155-623,Newton2013_ApJL779-L4}. The strength of neutron $^3P_2$ PBF process is stronger than that of proton $^1S_0$ PBF process, the neutron pairing has a larger influence on the NS cooling. In this work, we take neutron $^1S_0$ critical temperature from \cite{Wambach1993_NPA555-128}, Proton $^1S_0$ critical temperature from \cite{Amundsen1985_NPA437-487}, and neutron $^3P_2$ critical temperature $T_{\mathrm{cn}}^a$ from the Fig. 10 (curve ``a'') in \cite{Page2004_ApJSupp155-623} with the maximum critical temperature $10^9$ K.
Hereafter we adopt the DD-ME2 as the representative EoSs for most calculations.
Fig. \ref{fig:DD-ME2_Tc_profile}(a) illustrates the dependence of the critical temperatures of neutrons and protons on density. In the core of a NS (above the core-crust transition density), neutron pairing is predominantly in the $^3P_2$ channel. Above the onset density of $\Lambda$ hyperons, proton $^1S_0$ superfluidity can be neglected. Fig. \ref{fig:DD-ME2_Tc_profile}(b) shows that including $\Lambda$ hyperons alters the composition of NS matter. At high density, $\Lambda$ hyperons suppresses the neutron fraction, reducing the neutron Fermi momentum and resulting in an elevated neutron $^3P_2$ critical temperature. This issue arises because the neutron $^3P_2$ pairing model used here depends solely on the neutron Fermi momentum. A more self-consistent calculation of the neutron $^3P_2$ critical temperature and its application to NS cooling will be addressed in future work.

In our work, to systematically investigate the effects of the strength of superfluidity on the thermal relaxation of NSs, we fix the neutron and proton $^1S_0$ critical temperatures, while monotonically changing the neutron $^3P_2$ critical temperature through a re-scaling coefficient $R_{^3P_2}$, $T_{\mathrm{cn}}(k)=R_{^3P_2}T_{\mathrm{cn}}^a(k)$.

\section{The Thermal Relaxation and Cooling Simulation}\label{sec:results}
The thermal coupling between different structures during NS cooling is marked by a rapid decrease in surface temperature $T_s$ caused by the arrival of the cold front at the surface. Differences in the internal thermal structures of the NSs, e.g., the crust and core, the core region at which the dUrca processes are activated and the remaining core region, can lead to few separation cold fronts emerging the surface at different NS ages, further resulting in multiple rapid cooling regions on the cooling curve. Strong neutrino emission mechanisms, e.g., the dUrca processes, make the regions where they are activated too cold as a result of rapid cooling. This results in a strong heat flow directed toward these regions, ultimately disrupting the thermal coupling between other internal structures of the NS. Following the definition of \cite{Gnedin2001_MNRAS324-725} and \cite{Lattimer1994_ApJ425-802} , the thermal relaxation time is determined by
\begin{equation}\label{equ:tw}
    t_w=t~\text{for max}\left|\frac{\mathrm{d}\ln T_s}{\mathrm{d}\ln t}\right|,
\end{equation}
where $T_s$ is the surface temperature of NS and $t$ is the NS age. Generally speaking, the relaxation times are typically $t_w=$10--100 years, depending on the cooling models. $t_{\rm{w}}$ can reasonably approximated by $t_{\rm{w}}\approx\alpha t_1$ \citep{Lattimer1994_ApJ425-802,Gnedin2001_MNRAS324-725}, $\alpha$ is expressed as
\begin{equation}\label{equ:alpha}
    \alpha=\left(\frac{\Delta R_{\mathrm{crust}}}{1 \mathrm{km}}\right)^2\mathrm{e}^{-3\Phi}
\end{equation}
where $\Delta R_{\mathrm{crust}}$ is the crust thickness, $\mathrm{e}^{\Phi}=(1-2M/R)^{1/2}$. $t_1$ is the normalized relaxation time which depends solely on the microscopic properties of matter.

\begin{figure}
\centering
\includegraphics[width=1.0\linewidth]{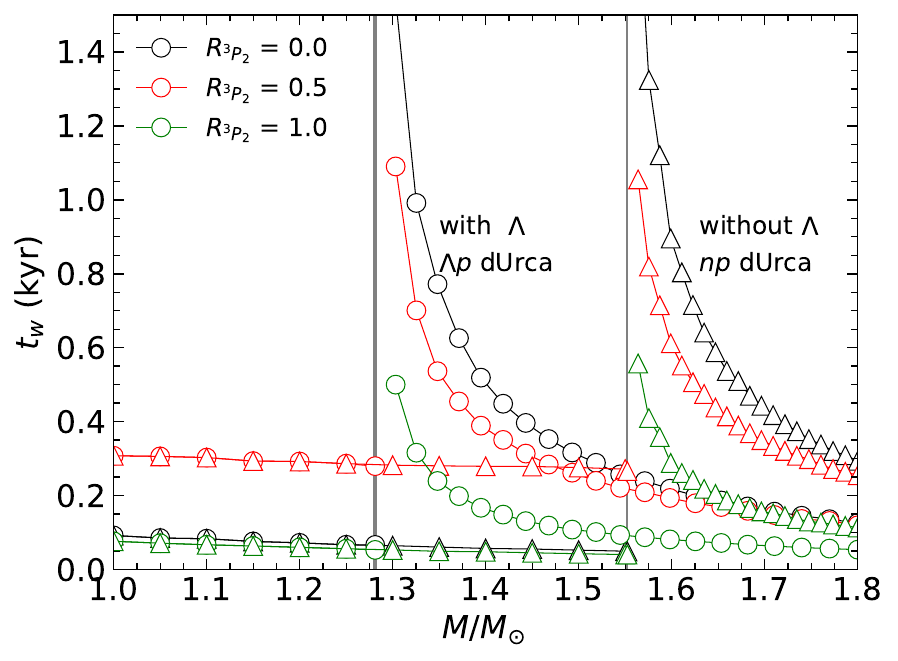}
\caption{\label{fig:DD-ME2L80_tw}Thermal relaxation time as a function of NS mass for DD-ME2 with $L_0=80$ MeV in the cases of with and without $\Lambda$ hyperons. The black, red, and green represent the relaxation times correspond to $R_{^3P_2}$=0.0, 0.5, 1.0 respectively. The hollow circle and upper triangle stand for the relaxation times calculated with EoSs with or without $\Lambda$ hyperons, respectively. For case with $\Lambda$ hyperons, the threshold mass at which the $\Lambda p$ dUrca process is activated is indicated by the black vertical dashed line ; while, for case without $\Lambda$ hyperons, the threshold mass at which the $np$ dUrca process is activated is shown by the gray vertical dashed line.
}
\end{figure}

With the microscopic inputs we described in the last section, we perform the cooling simulations of the NS by using the \texttt{NSCool} ~\footnote{http://www.astroscu.unam.mx/neutrones/
NSCool/} code. For different neutron $^3P_2$ critical temperatures, the dependence of the thermal relaxation time on the NS mass in the cases of with and without $\Lambda$ hyperons are given in Fig. \ref{fig:DD-ME2L80_tw}. We can observe that under finite physical conditions, the thermal relaxation of the star is delayed, with the thermal relaxation time being significantly larger than the typical value. These delayed thermal relaxations require the following physical conditions to be satisfied. One is that the presence of neutron $^3P_2$ superfluidity with a relatively low critical temperature. As shown in the Fig. \ref{fig:DD-ME2L80_tw}, the relaxation time corresponding to $R_{^3P_2}=0.5$ is longer than that at $R_{^3P_2}=1.0$. The other one is that, both for NSs with or without hyperon, the stellar mass need exceed the threshold mass at which the dUrca process is activated. When the stellar mass is just above the threshold mass, the thermal relaxation time is longer. The first condition involves the breaking and re-establishment of thermal coupling between the crust and core after the neutron $^3P_2$ PBF process is triggered. The second condition reflects the slow thermal relaxation between the activated core region (dU core) of the dUrca process and the remaining core region.

\subsection{Delayed Thermal Relaxation Triggered by Superfluidity}\label{sec:relaxSF}

\begin{figure}
\centering
\includegraphics[width=1.00\linewidth]{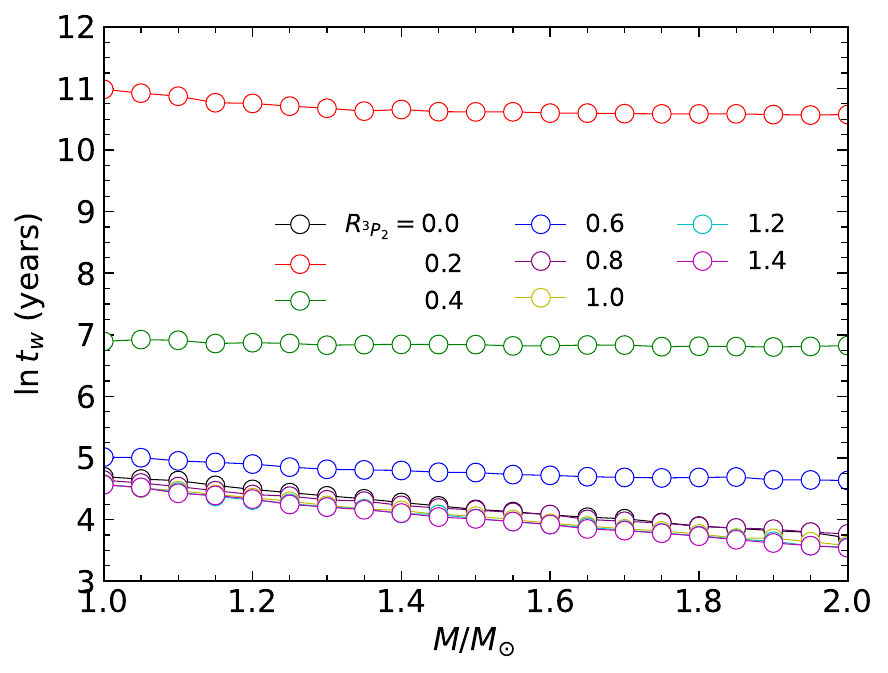}
\caption{\label{fig:DD-ME2L51_3P2}The thermal relaxation time as a function of NS mass for the original DD-ME2 with $L_0=51.26$ MeV. The $R_{^3P_2}$ ranges from 0.0 to 1.4.
}
\end{figure}

In Fig. \ref{fig:DD-ME2L51_3P2}, we show the thermal relaxation time as a function of NS mass with varying neutron $^3P_2$ critical temperature. We can see that the thermal relaxation time is very large for a low $T_{\rm{cn}}$; $t_w$ decreases as $T_{\rm{cn}}$ increases; finally, $t_w$ reach a typical value, which is dependent on NS mass, when the superfluidity strength is enough strong. These results can be explained by the trigger of nucleon $^3P_2$ PBF process. Generally speaking, the enhance cooling caused by PBF process is implemented when the internal temperature fall below $T_{\rm{cn}}$. If $T_{\rm{cn}}$ is small, in the early stage of NS cooling, the thermal coupling between crust and core is completed independently; after a waiting time $t_{\mathrm{wait}}$, the crust-core thermal coupling is broken after PBF process is triggered due to the strong PBF neutrino emission in the core; then new crust-core thermal coupling is reached after a new thermal relaxation time $t_{\rm{w}}^{\rm{PBF}}$. If $T_{\rm{cn}}$ is large, the PBF process is triggered in the early stage of NS cooling and hence the hybrid thermal relaxation close to original thermal relaxation without PBF process. We can approximate the total relaxation time by $t_{\rm{w}}\approx t_{\rm{wait}}+t_{\rm{w}}^{\rm{PBF}}$.

 Here we propose a simple analytic expression to fit the simulated $t_w$.
For the waiting time, if we assume the core to be isothermal few years after birth, then the global thermal balance gives
\begin{equation}\label{equ:balance1}
    C_V\frac{dT}{dt}=-R_{\rm{eff}}Q_\nu^{\rm{mU}},
\end{equation}
where $C_V$ is the total specific heat, $C_V=C_9T_9$ with $C_9\approx10^{39}~\rm{erg/K}$. $Q_\nu^{\rm{mU}}$ is the total neutrino emissivity of mUrca processes, $Q_\nu^{\rm{mU}}=Q_9T_9^8$ with $Q_9\approx10^{40}~\rm{erg/s}$ \citep{Page2011_PRL106-081101}.  $T_9=T/10^9$ K. If the pairing is considered, both the neutrino emissivity from mUrca processes and specific heat are suppressed, we introduce an effect global suppression factor $R_{\rm{eff}}$ to rescale total neutrino emissivity with $Q_\nu^{\rm{mU}}$ as the reference. The solution of Eq. \ref{equ:balance1} is
\begin{equation}\label{equ:waittime}
    t(T) = \tau_{\rm{mU}}^{\rm{eff}}\left( \frac{1}{T_9^6}-\frac{1}{T_{0,9}^6}\right),
\end{equation}
where we discard the initial age ($\approx1$ year) and $T_{0,9}$ is the initial internal temperature ($T_{0,9}\approx1.6$). $\tau_{\rm{mU}}^{\rm{eff}}=\tau_{\rm{mU}}/R_{\rm{eff}}$ is the cooling timescale with $\tau_{\rm{mU}}=10^9C_9/6Q_9\approx1.0$ year \citep{Page2011_PRL106-081101}. The waiting time is $t_{\rm{wait}}=t(T_{\mathrm{cn}})$. $t_w^{\rm{PBF}}$ involves the same thermal structure of the crust and is approximated by $\alpha t_1$.

The thermal relaxation time is written as follows
\begin{equation}\label{equ:relaxtime_pbf}
    t_{\rm{w}} \approx \tau_{\rm{mU}}^{\rm{eff}}\left( \frac{1}{T_{\mathrm{cn},9}^6}-\frac{1}{T_{0,9}^6}\right)\mathrm{e}^{-\Phi}+\alpha t_1,
\end{equation}
where $\mathrm{e}^{-\Phi}$ accounts for the gravitational dilation of time intervals. In small $T_{\mathrm{cn}}$ situation, $t_{\rm{wait}}$ could be $10^3$--$10^5$ years while the crust-core relaxation time is just a few decades, this means $t_w\approx t_{\rm{wait}}$. In the case of larger $T_{\mathrm{cn}}$, $t_{\rm{wait}}$ is negligible and $t_w\approx t_{w}^{\rm{PBF}}$. In principle, $t_1$ should be a function of $T_{\rm{cn}}$ because the superfluidity suppresses the specific heat. In practice, the fitted $t_1$ is the value of $R_{^3P_2}\rightarrow\infty$. The fixed $t_1$ in Eq. (\ref{equ:relaxtime_pbf}) is acceptable because $t_w^{\rm{PBF}}\ll t_{\rm{wait}}$ for small $T_{\rm{cn}}$. We emphasize that the change of crust-core relaxation time does not affect the mechanisms responsible for the delayed thermal relaxation that we are concerned with.

\begin{figure}
\centering
\includegraphics[width=1.0\linewidth]{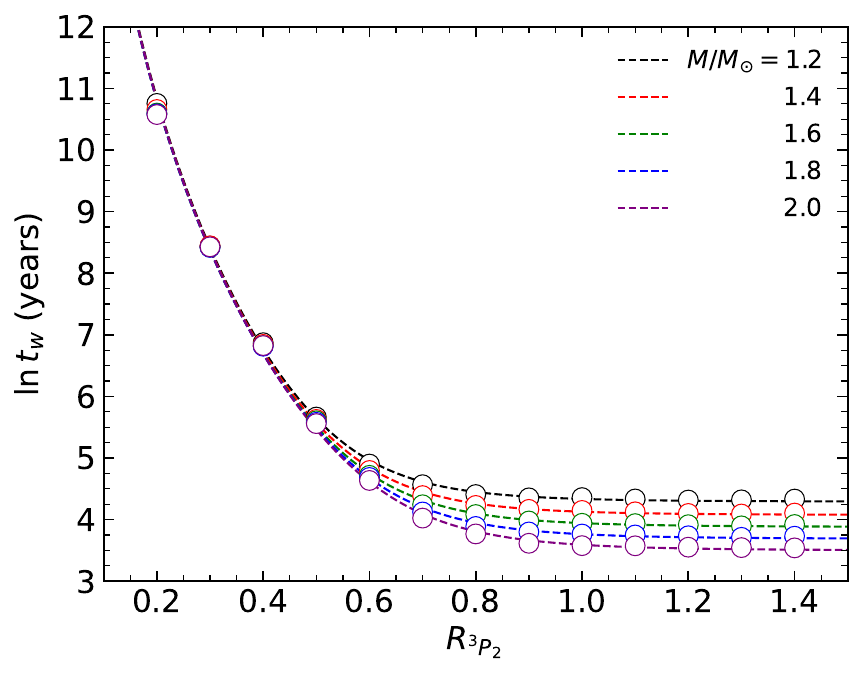}
\caption{\label{fig:DD-ME2L51_3P2_fitting}The thermal relaxation time as a function of $R_{^3p_2}$ simulated with original DD-ME2 with $L_0=51.26$ MeV for different stellar mass. Open circles are simulated relaxation time and the curves with the same color are the fitted relaxation time.}
\end{figure}

\begin{figure}
\centering
\includegraphics[width=1.0\linewidth]{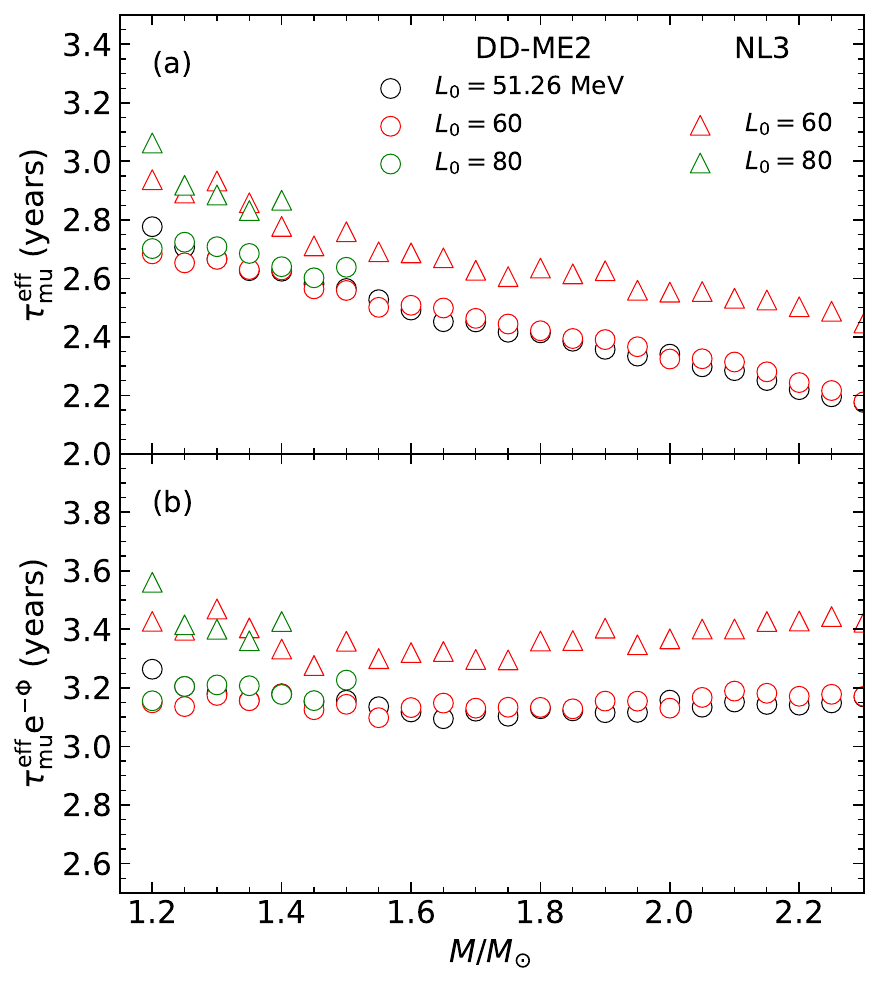}
\caption{\label{fig:tau}The fitted effective cooling timescale $\tau_{\mathrm{mU}}^{\mathrm{eff}}$ and $\tau_{\mathrm{mU}}^{\mathrm{eff}}\mathrm{e}^{\Phi}$ as a function of the stellar mass.
The calculations are done in the case of unified DD-ME2 and NL3 EoS models for different choices of symmetry energy slope $L_0$.
Closed and open circles represent $\tau_{\mathrm{mU}}^{\mathrm{eff}}$ and $\tau_{\mathrm{mU}}^{\mathrm{eff}}\mathrm{e}^{-\Phi}$, respectively. The fittings are performed for these stellar mass below the threshold mass $M_{c,N}^{np}$.}
\end{figure}

\begin{figure}
\centering
\includegraphics[width=1.0\linewidth]{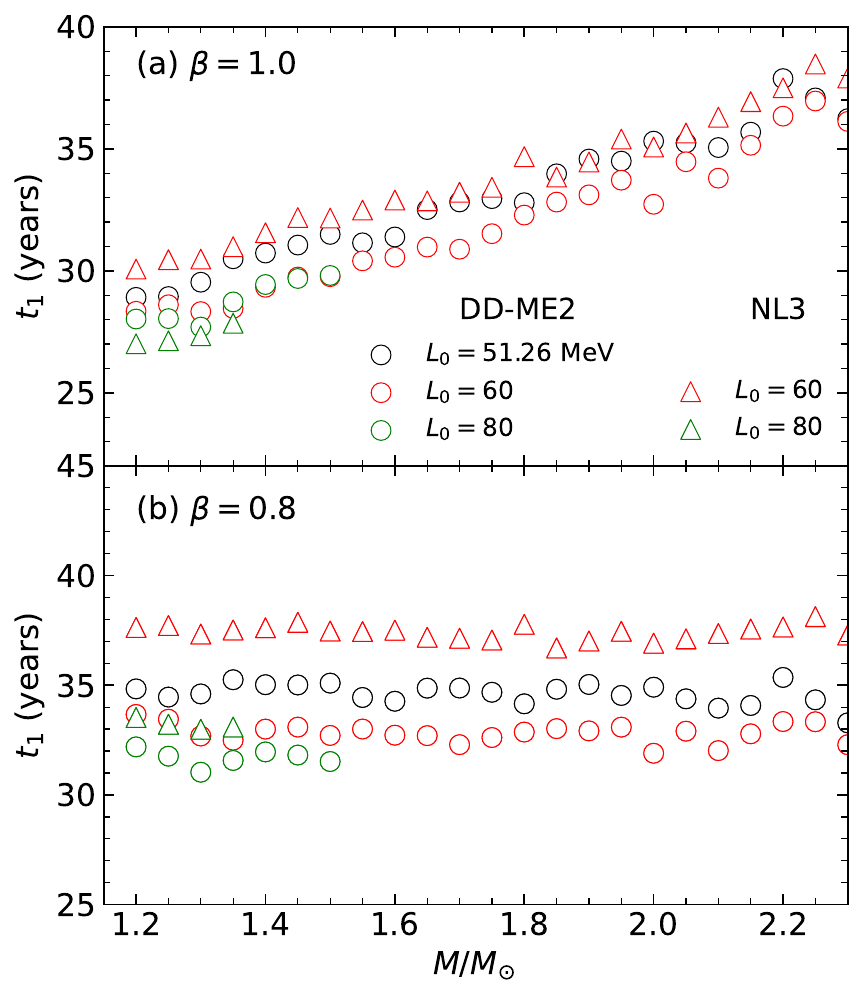}
\caption{\label{fig:t1}The fitted $t_1$ as a function of the stellar mass for (a) $\beta=1.0$ and (b) $\beta=0.8$.
The calculations are done in the case of unified DD-ME2 and NL3 EoS models for different choices of symmetry energy slope $L_0$.
The fittings are performed for the stellar masses below the threshold mass $M_{c,N}^{np}$.}
\end{figure}

\begin{table*}%[htbp]
  \centering
  \caption{The fitted cooling timescale $\tau_{\mathrm{mU}}^{\mathrm{eff}}\mathrm{e}^{-\Phi}$ and normalized relaxation time $t_1$.
  }\label{tab:taut1}
  \setlength\tabcolsep{14pt}
  \begin{tabular}{lrrrrrr}
    \hline
    \hline
     & \multicolumn{3}{c}{DD-ME2} & \multicolumn{3}{c}{NL3} \\
     \cmidrule(r){2-4} \cmidrule(r){5-7}
     $L_0$ (MeV) & 51.26 & 60 & 80 & 60 & 80 & 118.5 \\
    \hline
    $\tau_{\mathrm{mu}}^{\mathrm{eff}}\mathrm{e}^{-\Phi}$ (year) & 3.15$\pm$0.04 & 3.15$\pm$0.02 & 3.19$\pm$0.03 & 3.37$\pm$0.05 & 3.43$\pm$0.09 & -- \\
    $t_1(\beta=0.8)$ (year) & 34.62$\pm$0.48 & 32.82$\pm$0.44 & 31.69$\pm$0.37 & 37.41$\pm$0.33 & 33.21$\pm$0.23 & -- \\
    \hline
  \end{tabular}
\end{table*}

The fitted curves for $M/M_\odot=$1.0--2.0 are displayed in Fig. \ref{fig:DD-ME2L51_3P2_fitting}. We see that Eq. (\ref{equ:relaxtime_pbf}) can produce an excellent fit to the simulated thermal relaxation times. The fitted effective cooling timescale $\tau_{\mathrm{mu}}^{\mathrm{eff}}$ for the stellar mass below the threshold mass $M_{c,N}^{np}$ are shown in Fig. \ref{fig:tau} (a). $\tau_{\mathrm{mu}}^{\mathrm{eff}}$ exhibits a weak dependence on $L_0$ but shows a significant difference between effective interactions in the isospin scalar channel for DD-ME2 and NL3.
$\tau_{\mathrm{mu}}^{\mathrm{eff}}$ decreases with increasing stellar mass, this can be explained by the increasing core region dominated by mUrca processes. From Fig. \ref{fig:DD-ME2_Tc_profile}, before the neutron $^3P_2$ PBF process is triggered, the neutron and proton $^1S_0$ superfluidity suppress the neutrino emissivity of mUrca processes; when the stellar mass exceeds 1.2 $M_{\odot}$, the NS core contains an region that mUrca processes are not suppressed due to very weak neutron and proton $^1S_0$ superfluidity. This region expands as the stellar mass increases, leading to a faster cooling and a smaller cooling timescale. In Fig. \ref{fig:tau} (b), we find that the quantity $\tau_{\mathrm{mu}}^{\mathrm{eff}}\mathrm{e}^{-\Phi}$ have no obvious dependence on the stellar mass and therefor $t_{\rm{wait}}$ is only sensitive to EoSs and $T_{\mathrm{cn}}$. The fitted $t_1$ are given in Fig. \ref{fig:t1}(a), $t_1$ increases as the stellar mass increases. Following the definition in \cite{Lattimer1994_ApJ425-802}, we expect that $t_1$ depends solely on the microscopic properties of matter instead of the stellar properties. We rewritten $t_{w}^{\rm{PBF}}$ as $\alpha^{\beta}t_1$. \cite{Lattimer1994_ApJ425-802} have found that $\beta\neq 1$ for different choices of the crust-core transition density and superfluidity strength. We find that, for $\beta\approx0.8$, $t_1$ have no obvious dependence on the macroscopic properties of the star, as shown in Fig. \ref{fig:t1}(b). We list $\tau_{\mathrm{mU}}^{\mathrm{eff}}\mathrm{e}^{-\Phi}$ and $t_1(\beta=0.8)$ for different effective interactions in Table \ref{tab:taut1}.

We cannot observe the delayed thermal relaxation if $T_{\rm{cn}}$ is too small. In the photon emission dominated cooling stage (age $>10^5$ years), the rapid cooling driven by PBF process is hidden by photon emission. If we set that the photon emission dominated cooling starts at $\sim10^5$ years, we find that observable delayed thermal relaxation requires $T_{\mathrm{cn}}>0.18\times10^9$ K. When $T_{\mathrm{cn}}<0.18\times10^9$ K, we can only observe the thermal relaxation with a typical relaxation time.

Note that $T_{\rm{cn}}$ used in Eqs. (\ref{equ:waittime}) and (\ref{equ:relaxtime_pbf}) is the actual maximum value of neutron $^3P_2$ critical temperature inside NSs, not the maximum value of $R_{^3P_2}T_{\mathrm{cn}}^a(k)$. In our all simulations, for the neutron $^3P_2$ critical temperature, the Fermi momentum corresponds to the maximum value of $R_{^3P_2}T_{\mathrm{cn}}^a(k)$ can be satisfied even for the stars with $1.0M_{\odot}$. This allows us to utilize observations to set constraints on the theoretical maximum $T_{\rm{cn}}$ inside a NS. Cassiopeia A (Cas A) is a candidate of delayed thermal relaxation NS, undergoing a rapid drop in surface temperature of $2\%$–$5.5\%$ at the age of 335 years \citep{Newton2013_ApJL779-L4}. Using $\tau_{\rm{mU}}^{\rm{eff}}\mathrm{e}^{-\Phi}\approx3.2$ years and $\alpha t_1\approx60$ years, we estimate the neutron $^3P_2$ critical temperature to be $T_{\rm{cn}}\approx0.47\times10^9$ K.
Recently, \citet{Alford2024_PRC110-L052801} proposed a systematically improvable approach to the Urca rate calculation by applying the nucleon width approximation, they found an enhancement of the mUrca rate by more than an order of magnitude. To study the corresponding effect, we simply increase the neutrino emissivity of mUrca processes to its 10 times. The suppression of superfluidity on the local neutrino emissivity is decoupled $Q_\nu=R(T/T_c)Q_\nu^{\rm{mU}}(\rho,T)$, we reasonably estimate $R_{\rm{eff}}Q_\nu^{\rm{mU}}\rightarrow10R_{\rm{eff}}Q_\nu^{\rm{mU}}$ in Eq. \ref{equ:balance1}. This results in a reduction of the cooling timescale to 0.32 years and the neutron $^3P_2$ critical temperature $T_{\rm{cn}}$ to $0.32\times10^9$ K.
Note that here we are only analyzing the rapid cooling of Cas A from the perspective of the delayed thermal relaxation. The rapid cooling of Cas A requires a surface temperature drop of $2\%$–$5.5\%$, which may involve more physics, e.g., proton superconductivity, the density dependence of critical temperatures, and envelope models. Relevant discussions on these topics will be addressed in our future work.

\subsection{Delayed Thermal Relaxation Triggered by dUrca Processes}\label{sec:relaxdUrca}%%%%%%%%%%
For both NSs with and without hyperons, delayed thermal relaxation can always be observed above the threshold mass at which dUrca processes are activated.
In the present work, considering that $L_0$ is constrained to be below 60 MeV \citep{Hooker2013_JPCS420-012153,Newton2013_ApJL779-L4,Tu2024_arXiv2412.09219}, the $np$ dUrca process is prohibited. Therefore, in our work, we will focus on the delayed thermal relaxation caused by the $\Lambda p$ dUrca process.

\begin{figure}
\centering
\includegraphics[width=1.0\linewidth]{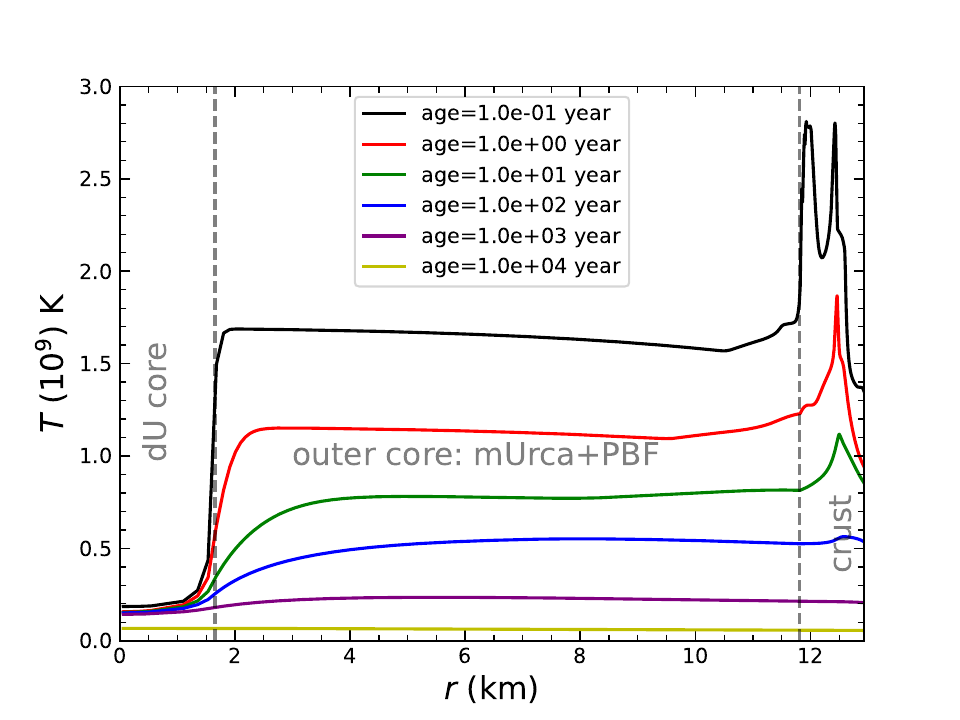}
\caption{\label{fig:TempProfile}The internal temperature profiles in the $M=1.3299M_\odot$ NS. The adopted EoS is the original DD-ME2 (with $L_0=51.26$ MeV) and $R_{^3P_2}=0.0$. Outer-inner core and crust-core interfaces (left and right black dashed vertical lines) divide the NS interior into three parts: dU allowed kernel, outer core, and crust. In the dU allowed kernel, the hyperon dU process $\Lambda p$ channel is opened.}
\end{figure}

In Fig. \ref{fig:TempProfile}, we exhibit the internal temperature profiles in the $M=1.3299M_\odot$ NS. The outer-inner core interface divides the core into dU core and outer core, the dUrca process dominates the fast cooling in the former while mUrca and PBF processes dominate the standard cooling in the later. The colder dU core gains heat from the heat flows from the warmer outer core due to the very large temperature derivative in the interface, so the temperature $T_{\rm{dU}}$ of dU core remains almost constant until $\sim1000$ year old, see Fig. \ref{fig:TempProfile}.
The heat compensation to dU core results in the faster cooling in the outer core, this could affect the crust-core thermal relaxation, or say the lower peaks in Figs. 6-9 of \cite{Sales2020_A&A642-A42}. If the temperature of outer core decreases to a characteristic temperature $T_{\rm{t}}$, then dU core and outer core complete their thermal coupling and act as a core. The thermal relaxation between the new core and crust corresponds to the larger peaks in Figs. 6-9 of \cite{Sales2020_A&A642-A42}.
When the dU core is enough large, the rapidly drop in the  temperature of outer core leads to a fast thermal relaxation and short relaxation timescale, the relaxation time can be expressed by Eq. (5) in \cite{Gnedin2001_MNRAS324-725}.
Nevertheless, Eq. (5) in \cite{Gnedin2001_MNRAS324-725} cannot explain the delayed thermal relaxation we observed above $M_{c,\Lambda}^{\Lambda p}$.

Before addressing the origin of the delayed thermal relaxation in the core, it is worth noting the clear temperature fluctuations seen in the crust region, as shown in Fig.~\ref{fig:TempProfile}. Here, we briefly explain the underlying mechanisms; for more detailed discussion, see \citet{Gnedin2001_MNRAS324-725}.
Due to the higher neutrino emissivity in the core, the crust remains hotter during the early cooling phase of NSs (age < crust-core thermal coupling timescale).
Simultaneously, the crust exhibits significant temperature fluctuations; specifically, two distinct temperature peaks emerge in both the inner and outer crust regions. At the interface between the inner and outer crusts, neutrons drip out from the nuclei to form a superfluid and contribute to neutrino emissivity and specific heat. With increasing density, the dripped neutrons display stronger superfluidity and thus reduce the neutrino emissivity and specific heat; see \cite{Gnedin2001_MNRAS324-725} for more details.
This leads to two effects in the crustal region near the boundary: on the one hand, the enhanced neutrino emission from neutron Cooper pairs accelerates cooling in this region; on the other hand,  the increased specific heat due to dripped neutrons creates a ``heat barrier'', causing the heat transfer from the inner crust to the core to be significantly faster than that from the outer to inner crust.
Consequently, the temperature peak in the outer crust persists until crust-core thermal coupling is achieved. 

We then propose a simple model and deduce an analytical expression to explain the delayed thermal relaxation due to the following assumptions: the whole dU+outer core is isothermal; the temperature of dU core remains constant before it is coupled to outer core; both dU core and outer core have dependent global thermal evolution and the outer core is responsible for compensating energy loss of dU core; the neutrino emissivities of mUrca processes for different regions are proportional to their volumes. The second assumption is not strictly correct because we can see the slow decline in the dU core temperature, see Fig. \ref{fig:TempProfile}, the decline is more pronounced for the larger dU core; we think the assumption is reasonable when the dU core is small because the large heat capacity in the outer core can easily compensate the energy loss of the dU core. The fourth assumption neglect the component differences from different regions but we can see that the assumption has captured the main feature in the thermal evolution.

The model is described as follows. The temperature of outer core is evolved from the initial temperature $T_0$ at the initial time ($\sim1$ years), the temperature of dU core remains a constant. The energy balances are
\begin{widetext}
\begin{equation}\label{equ:balance_outercore}
    C_V\frac{dT}{dt}=-R_{\rm{eff}}Q_\nu^{\rm{mU}}\theta(T-T_{\rm{cn}})-f_{\rm{PBF}}Q_\nu^{\rm{mU}}\theta(T_{\rm{cn}}-T)-Sf_{\rm{dU}}Q_9T_{\rm{dU}}^8/f_V,
\end{equation}
\end{widetext}
for the outer core and
\begin{equation}\label{equ:balance_DU}
    C_V\frac{dT}{dt}=0,
\end{equation}
for the dU core. In Eq. \ref{equ:balance_outercore}, $f_{\rm{PBF}}$ is the ratio of the neutrino emissivity of PBF proesses to that of mUrca processes; because the thin neutrino emission spherical shells are proportional to $T$, after integrated over the core volume, $Q_\nu^{\rm{PBF}}$ can be reasonably approximated by a $T^8$ law \citep{Gusakov2004_A&A423-1063}, hence $Q_\nu^{\rm{PBF}}=f_{\rm{PBF}}Q_\nu^{\rm{mu}}$ with $f_{\rm{PBF}}\sim$10 \citep{Page2004_ApJSupp155-623,Gusakov2004_A&A423-1063}. The PBF processes are more efficient than the mUrca processes, we use the step function to neglect the neutrino emissivity of mUrca processes when $T<T_{\rm{cn}}$. The right third term of Eq. \ref{equ:balance_outercore} is a constant energy loss of dU core; $f_{\rm{dU}}$ is the ratio of the local neutrino emissivity of dUrca process to that of mUrca process, $f_{\rm{dU}}\sim5\times10^5T_9^{-2}$ \citep{Lattimer1991_PRL66-2701}; $f_{\rm{V}}$ is the ratio of the outer core volume $V_{\rm{out}}=4\pi(R_{\rm{core}}^3-R_{\rm{dU}}^3)/3$ to the dU core volume $V_{\rm{dU}}=4\pi R_{\rm{dU}}^3/3$, where $R_{\rm{core}}$ and $R_{\rm{dU}}$ are the star's core radius and dU core radius respectively; $S$ is the ratio of the neutrino emissivity of a dUrca process to that of $np$ dUrca process, $S\sim0.04$ for $\Lambda p$ channel \citep{Prakash1992_ApJ390-L77}.

The solution of Eq. \ref{equ:balance_outercore} at $T=T_{\rm{t}}$ is the relaxation time between the outer core and dU core, we get the correct relaxation time after the thermal coupling of crust and core,
\begin{widetext}
\begin{equation}\label{equ:relaxtime_DU}
    t_{w}\approx-\frac{6\tau_{\rm{mU}}\mathrm{e}^{-\Phi}}{R_{\rm{eff}}}\int_{T_0}^{T_{\rm{cn}}}\frac{T_{9}}{T_{9}^{8}+a_{1}}dT_{9}-\frac{6\tau_{\rm{mU}}\mathrm{e}^{-\Phi}}{f_{\rm{PBF}}}\int_{T_{\rm{cn}}}^{T_{\rm{t}}}\frac{T_{9}}{T_{9}^{8}+a_{2}}dT_{9}+\alpha t_{2},
\end{equation}
\end{widetext}
where two acceleration factors from the dU core are expressed by
\begin{equation}\label{equ:accelerationfactor}
    a_1 = \frac{Sf_{\rm{dU}} T_{\rm{dU},9}^8/R_{\rm{eff}}}{[(R_{\rm{core}}/R_{\rm{dU}})^3-1]}, \quad
    a_2 = \frac{Sf_{\rm{dU}} T_{\rm{dU},9}^8/f_{\rm{PBF}}}{[(R_{\rm{core}}/R_{\rm{dU}})^3-1]}.
\end{equation}
From Sec. \ref{sec:relaxSF} and Eq. \ref{equ:relaxtime_DU}, the information of EoS and NS structure are compiled into the acceleration factors, e.g., $T_{\rm{dU}}$ and $R_{\rm{dU}}$.

\begin{figure}
\centering
\includegraphics[width=1.00\linewidth]{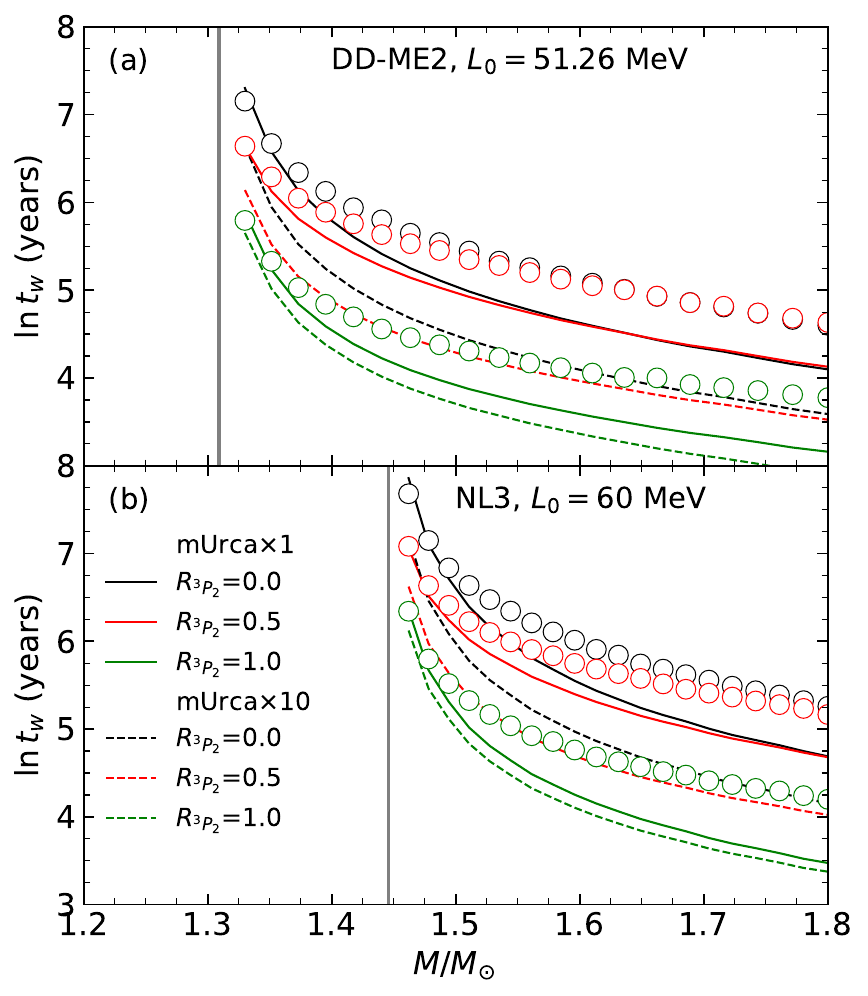}
\caption{\label{fig:fitting_tw_durca}The relaxation time as a function of NS mass above $M_{\rm{c,\Lambda}}^{\Lambda p}$ for (a) the original DD-ME2 with $L_0=51.26$ MeV and (b) NL3 with $L_0=60$ MeV. The dashed vertical lines represent the threshold mass $M_{\rm{c,\Lambda}}^{\Lambda p}$ of two EoSs, respectively. Open circles are simulated relaxation time and the curves with the same color are obtained by using Eq. \ref{equ:relaxtime_DU}. The dashed curves give the relaxation time when the neutrino emissivity of mUrca process is increased by a factor of 10.}
\end{figure}

We use the simulated results to validate our model. Fig. \ref{fig:fitting_tw_durca} demonstrates the simulated relaxation time and the relaxation time obtained by Eq. \ref{equ:relaxtime_DU} as a function of NS mass above $M_{\rm{c,\Lambda}}^{\Lambda p}$, taking the original DD-ME2 (with $L_0=51.26$ MeV) and NL3 with $L_0=60$ MeV as examples. We adopt $t_2\approx6$ years \citep{Gnedin2001_MNRAS324-725} for rapidly cooling in Eq. (\ref{equ:relaxtime_DU}). $T_{\rm{t}}\approx T_{\rm{dU}}$ is reasonable if the dU core is small while $T_{\rm{t}}$ should be larger than $T_{\rm{dU}}$ for keeping the thermal coupling between dU core and outer core. Besides, the larger dU core the smaller $T_{\rm{dU}}$. In Fig. \ref{fig:fitting_tw_durca}, we suppose $T_{\rm{t}}\approx T_{\rm{dU}}$ for all NS mass. We can see that the tendency of $t_w$ changes with NS mass is perfectly reproduced by Eq. (\ref{equ:relaxtime_DU}). For the original DD-ME2, Eq. (\ref{equ:relaxtime_DU}) match simulated relaxation time well for these NSs just above $M_{\rm{c,\Lambda}}^{\Lambda p}$ if $T_{\rm{t},9}\approx$ 0.150, 0.145, and 0.160 for $R_{^3P_2}=0.0,0.5,1.0$, respectively; the significant departure from Eq. (\ref{equ:relaxtime_DU}) can be found in the region of massive NSs, the reason is quite simple because the heavier NS, the lower $T_{\rm{dU}}$, the smaller acceleration factor, finally the larger relaxation time. We can obtain similar results for NL3 with $L_0=60$ MeV, $T_{\rm{t},9}\approx$ 0.145, 0.140, and 0.155 for $R_{^3P_2}=0.0,0.5,1.0$, respectively. The neutron $^3P_2$ PBF process accelerates the cooling of the outer core; as $T_{\mathrm{cn}}$ increases, the PBF process occurs earlier, such that larger $T_{\mathrm{cn}}$ correspond to shorter relaxation times, as we can see from Fig. \ref{fig:fitting_tw_durca}. All the qualitative results of the above analysis hold for other EoSs used in our work.

We also considered the enhancement of the mUrca rate as we done in Sec. \ref{sec:relaxSF} by setting $R_{\rm{eff}}Q_\nu^{\rm{mU}}\rightarrow10R_{\rm{eff}}Q_\nu^{\rm{mU}}$ in the first term of Eq. (\ref{equ:relaxtime_DU}). Because of $f_{\rm{PBF}}\sim10$, the neutrino emissivities of PBF and enhanced mUrca processes are the same order of magnitude. We simply halve the second term in Eq. (\ref{equ:relaxtime_DU}), or equivalently, set $f_{\rm{PBF}}\sim20$. The dUrca rate have no significant change \citep{Alford2024_PRC110-L052801}, we leave the third term of Eq. (\ref{equ:relaxtime_DU}) unchanged. From Fig. \ref{fig:fitting_tw_durca}, the enhanced mUrca rate significantly shortens the thermal relaxation time. When $T_{\mathrm{cn}}$ is small, the cooling of the outer core is primarily contributed by the mUrca processes; an order-of-magnitude increase in the neutrino emissivity is obtained from the enhanced mUrca processes, this leads to a prominent reduction in the thermal relaxation time. When $T_{\mathrm{cn}}$ is large, the cooling of the outer core is mainly contributed by both PBF and mUrca processes. Compared to the scenario with only the PBF process, there is no significant increase in the neutrino emissivity, and thus the shortening of the thermal relaxation time is less pronounced.

\section{Summary and Perspective}\label{sec:summary}
The thermal relaxation behavior for the cooling of NSs not only has the potential to constrain the EoS of dense matter, but also to probe the pairing properties of dense nuclear matter. In this work, we systematically investigate the thermal relaxation properties of rapid cooling NSs induced by the PBF and dUrca processes with several effective interactions in different isospin vector and scalar channels. We find that, under specific physical conditions, rapid cooling NSs exhibit delayed thermal relaxation phenomena.

On one hand, the delayed thermal relaxation caused by the PBF process involves the breaking and re-establishment of the thermal relaxation between the core and crust. For a low value of $T_{\mathrm{cn}}$, the stellar core needs a longer time to cool down to an internal temperature that enables the trigger of the PBF process, and therefore the small $T_{\mathrm{cn}}$ is required. We propose a simple model to describe this delayed thermal relaxation. and then we constrain the neutron $^3P_2$ critical temperature as $0.47\times10^9$ K and $0.32\times10^9$ K for standard and enhanced mUrca rates with the observations of Cas A.

On the other hand, the delayed thermal relaxation induced by the dUrca process arises from the slow thermal coupling between the small-size dU core and the outer core. Because the large-size dU core dramatically absorbs heat from the outer crust and accretes the thermal coupling with the outer core, the stellar mass is required above but close to the threshold mass for which the dUrca process is activated. A simple analytical model is also proposed, while it describes NSs with masses close to the threshold mass more accurately. The enhanced mUrca rate
can shorten the delayed relaxation time, but this is not noticeable when $T_{\rm cn}$ is large enough.

For future work, we plan to develop the proposed analytical formulas further to more clearly connect key physical quantities, such as $L_0$ and $T_{\mathrm{cn}}$, to observational data.
Additionally, the evolution of magnetic field may alter the electron conductivity and other transport properties. we also aim to conduct two- or three-dimensional simulations for the thermal relaxation of NSs, incorporating magnetic field evolution.

\section*{Acknowledgments}
We thank Prof. R. Negreiros for valuable discussions and suggestions.
The work is supported by the National SKA Program of China (No.~2020SKA0120300) and the National Natural Science Foundation of China (grant Nos.~12273028 and 12494572).

\bibliographystyle{aasjournal}%{unsrt}
\bibliography{ref}

\end{document}